\documentclass{PoS}

\title{S-wave meson-baryon potentials with strangeness from Lattice QCD}

\ShortTitle{S-wave meson-baryon potentials with strangeness from Lattice QCD}

\author{\speaker{Yoichi Ikeda}\\
        Department of Physics, Tokyo Institute of Technology, 
        Meguro, Tokyo 152-8551, Japan\\
        E-mail: \email{yikeda@riken.jp}}
\author{for HAL QCD Collaboration\\ 
\includegraphics[width=.40\textwidth]{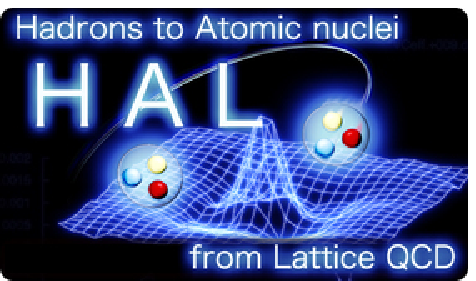}}

\abstract{
We study 
the s-wave $I=2$ $\pi \Sigma$ and $I=1$ $KN$ interactions
from 2+1 flavor full lattice QCD simulation for relatively
heavy quark mass corresponding to $m_{\pi}=700$ MeV.
The s-wave meson-baryon potentials are obtained from the Nambu-Bethe-Salpeter
amplitudes.
Potentials in both channels reveal short range repulsions, 
which suggest the importance of the Pauli blocking effect.
The $I=1$ $KN$ scattering phase shifts are calculated
and compared with the existing experimental data. 
          }

\FullConference{ The XXIX International Symposium on Lattice Field Theory - Lattice 2011\\
July 10-16, 2011\\
Squaw Valley, Lake Tahoe, California}

\begin{document}

\section{Introduction}

The $\Lambda(1405)$ negative parity hyperon resonance 
has strangeness $S=-1$ and isospin $I=0$, 
and is located just below the $\bar KN$ threshold.
Since the $\Lambda(1405)$ is considered to be the quasi-bound state 
of the s-wave $\bar{K}N$, 
the structure of the $\Lambda(1405)$ is one of the very important issues 
of recent hadron physics, 
especially to understand $\bar K$-nucleon and $\bar K$-nucleus interactions. 
The $\Lambda(1405)$ also decays into the $\pi \Sigma$ continuum.
Thus, for the physics of the $\Lambda(1405)$ and the $\bar K$-nucleus, 
dynamics of both $\pi\Sigma$ and $\bar KN$ is important 
and gives essential contributions.

The $\Lambda(1405)$ has been considered as a dynamically generated state 
in meson-baryon scattering 
and well described in a coupled-channels approach 
based on chiral dynamics~\cite{chiral}. 
The chiral dynamics also predicts two resonance poles for the $\Lambda(1405)$,
which have different coupling nature 
to the $\pi\Sigma$ and $\bar KN$ channels~\cite{Jido:2003cb}.
A phenomenological approach also described 
the $\Lambda(1405)$ as a quasi-bound state of $\bar KN$~\cite{AY02}, 
in which an effective interaction of $\bar KN$ was derived so as to 
reproduce the $\bar KN$ scattering length
and the mass and width of the $\Lambda(1405)$ as 1405 MeV and 40 MeV.
With this approach, only one resonance pole for the $\Lambda(1405)$ is predicted,
and the phenomenological model provides quantitatively 
stronger $\bar KN$ interaction 
than the chiral potential in the region far below the $\bar{K}N$ threshold.
Thus, there are uncertainties about the theoretical extrapolation 
of the $\bar K N$ interaction below the $\bar K N$ threshold and the
pole nature of the $\Lambda(1405)$.

Recently, threshold behavior of the $\pi\Sigma$ scattering
and its impact on the binding energy of the $\bar K$-nuclei
have been discussed~\cite{Ikeda_pS}.
For the physics of the $\Lambda(1405)$, it is certainly necessary to 
have theoretical descriptions of $\pi\Sigma$ and $\bar KN$ dynamics. 
The position of the pole singularity 
(bound state, virtual state, or resonance) around the $\pi\Sigma$ threshold
is an important issue to investigate the $\bar K$-nuclei~\cite{AY02,KNN}.
The $\pi \Sigma$ dynamics in $I=0$ channel can be extracted 
from the $\Lambda_c$ baryon decay to $\pi \pi \Sigma$ states \cite{HO:2011}.
According to Ref~\cite{HO:2011}, in the $\Lambda_c$ decay process,
it is possible to provide two constraints 
on the three different isospin component of the $\pi \Sigma$ scattering lengths.
Therefore, the direct determination of 
one of $\pi \Sigma$ scattering lengths based on QCD is mandatory.

In this paper,
to clarify the nature of the $\pi\Sigma$ dynamics, 
we calculate the $I=2$ $\pi \Sigma$ potential on the lattice.
The method to calculate the potentials 
from Nambu-Bethe-Salpeter wave function, which satisfies
the relativistic three-dimensional Schr\"{o}dinger-type equation,
has been reported in Refs.~\cite{IAH:06,AHI:09},
and is developed by HAL QCD Collaboration~\cite{
Nemura2008,Inoue2010a,Ikeda2010,Sasaki2010,Murano2011,Inoue2010b,Doi2011,Aoki_HAL}.
Further applications are also given in Refs.~\cite{
Kawanai2010,TTT2009,QQ}.
We apply this method to the $I=2$ $\pi \Sigma$ system.
We also examine the $I=1$ $KN$ scattering,
since the $I=1$ $KN$ belongs to the same multiplet 
as the $I=2$ $\pi \Sigma$ in flavor SU(3) limit, and
there exist experimental data in this channel.
Therefore, examining the $I=1$ $KN$ scattering is very useful to
predict the $I=2$ $\pi \Sigma$ scattering observable.

This paper is organized as follows.
In section 2, the formalism to extract the meson-baryon potentials from lattice QCD
is briefly reviewed.
Our numerical setup of the lattice QCD simulation is then shown in section 3,
and the results are shown and discussed in section 4.
A summary is given in section 5.

\section{Formalism}
\subsection{Basic concept to define potentials on the lattice}
Following the basic formulation to extract the nucleon-nucleon interaction~\cite{IAH:06,AHI:09},
we briefly show our strategy to obtain meson-baryon potentials below.
We start with the Schr\"{o}dinger-type equation for 
the equal-time Nambu-Bethe-Salpeter (NBS) wave function $\Psi_{\vec k}(\vec{r})$:
\begin{equation}
( \nabla^2 + \vec k^2 ) \psi_{\vec k}(\vec{r}) 
= 2 \mu \int d\vec{r'} U(\vec{r},\vec{r'}) \psi_{\vec k}(\vec{r'})~,
\label{Schrodinger-type1}
\end{equation}
where $\mu(=mM /(m + M))$ and $\vec k$ denote the reduced mass of 
the mason ($m$) and the baryon ($M$)
and the relative momentum of the meson-baryon system, respectively.
The NBS wave function of the meson-baryon system is extracted from
the four-point correlation functions on the lattice:
\begin{eqnarray}
C_{\alpha}(\vec r, t-t_0) &=&
\sum_{\vec x, \vec X, \vec Y}
\left\langle 0 \right|
\phi_{M}(\vec x+\vec r, t) \phi_{B,\alpha}(\vec x, t)
\bigl(
P^{(s)}_{\beta}
\phi_{M}(\vec X, t_0)  \phi_{B,\beta}(\vec Y,t_0)
\bigr)^{\dagger}
\left| 0 \right\rangle \nonumber \\
& = &
\sum_{n, \vec x} A_{n} \left\langle 0 \right|
\phi_{M}(\vec x+\vec r, t) \phi_{B,\alpha}(\vec x, t)
\left| n \right\rangle 
\ e^{-E_n(t-t_0)}~,
\label{4-point}
\end{eqnarray}
with the spin projection operator $P^{(s)}$ and the matrix elements
\begin{equation}
A_{n} = 
\sum_{\vec X, \vec Y}
\left\langle n \right| 
\bigl(
P^{(s)}_{\beta} 
\phi_{M}(\vec X, t_0)  \phi_{B,\beta}(\vec Y,t_0)
\bigr)^{\dagger}
\left| 0 \right\rangle~.
\end{equation}
The four-point correlation function in Eq. (\ref{4-point})
is dominated by the lowest energy state with total energy $E_0$
at large time separation ($t \gg t_0 $):
\begin{eqnarray}
C_{\alpha}(\vec r, t-t_0)
&\rightarrow&
A_0 \psi_{\alpha}(\vec r ~; J^{\pi}) e^{-E_0(t-t_0)}~,
\label{4-point2}
\end{eqnarray}
with $E_0=\sqrt{m^2+\vec k^2}+\sqrt{M^2+\vec k^2}$ being the relativistic energy 
of the meson-baryon system.
Thus, the meson-baryon NBS wave function is defined by the spatial correlation of
the four-point correlation function.
In Eq. (\ref{4-point2}), we assume the Dirichlet boundary condition
in temporal direction, so that the temporal correlation has 
an expnential form, $e^{-E_0(t-t_0)}$.

The NBS wave function in s-wave state is obtained under the projection onto 
the $A_1^+$ sector,
\begin{equation}
\psi(\vec r ~; J^{\pi}=1/2^-) =
\frac{1}{24}\sum_{g \in O} P_{\alpha} 
\psi_{\alpha}(g^{-1} \vec r ~; J^{\pi})~,
\label{BS-wave}
\end{equation}
where $g \in O$ represent 24 elements of the cubic rotational group,
and the summation is taken for all these elements.
Using Eq. (\ref{Schrodinger-type1}) and Eq. ({\ref{BS-wave}}),
we will find the meson-baryon potential and wave function from lattice QCD.


The energy-independent and non-local potential $U(\vec{r},\vec{r'})$ can be expanded
in powers of the relative velocity $\vec{v}=-i\nabla/\mu$ at low energies,
\begin{eqnarray}
U(\vec{r},\vec{r'}) & = &
V(\vec{r}, \vec{v}) \delta(\vec{r} - \vec{r'})  \nonumber \\
& = & (V_{LO}(\vec{r}) 
+ (\vec L \cdot \vec{\sigma}) V_{NLO}(\vec{r})
+ \cdots) \delta(\vec{r}-\vec{r'})~,
\label{velocity}
\end{eqnarray}
where the $N^n LO$ term is of order $O(\vec{v}^n)$,
and 
$\vec L$ and $\vec{\sigma}$ denote an orbital angular momentum of
the meson-baryon system and a baryon spin, respectively.


\subsection{Strategy to extract potentials on the lattice}
In lattice simulations, one may suffer from possible contaminations from
higher energy states in Eq. (\ref{4-point2}).
To extract the reliable potentials on the lattice~\cite{Ishii_t}
even with the presence of such higher energy states,
we consider the time evolution 
of normalized four-point correlation functions (R-correlators)~\cite{Ishii_t}:
\begin{eqnarray}
-\frac{\partial}{\partial t} R_{\alpha}(\vec r, t-t_0) 
&=&
\sum_{n}A_{n} \Delta E(\vec k_n) 
e^{- \Delta E(\vec k_n) (t-t_0)} \psi_{\alpha, \vec k_n}(\vec r) \nonumber \\
&\simeq&
\sum_{n}A_{n} 
\Biggl(
\frac{\vec k_n^2}{2\mu} + U
\Biggr)
e^{- \Delta E(\vec k_n) (t-t_0)} \psi_{\alpha, \vec k_n}(\vec r)~,
\label{R-eq}
\end{eqnarray}
with
\begin{eqnarray}
R_{\alpha}(\vec r, t-t_0) &\equiv& e^{(m+M)(t-t_0)}C_{\alpha}(\vec r, t-t_0)~, \\
\Delta E(\vec k_n) &=& \sqrt{m^2+\vec k_n^2} + \sqrt{M^2 + \vec k_n^2} - (m+M)~.
\end{eqnarray}
In Eq. (\ref{R-eq}), we assume the non-relativistic energy levels.
Since the potential $U(\vec r, \vec r')$ is non-local but energy-independent 
by construction~\cite{IAH:06,AHI:09},
one can define the potentials even with
the higher energy state~\cite{Ishii_t}.
At the leading order of the velocity expansion in Eq. (\ref{velocity}) and
after the s-wave projection through Eq. (\ref{BS-wave}),
Eq. (\ref{R-eq}) reads
\begin{eqnarray}
-\frac{\partial}{\partial t} R(\vec r, t-t_0 ~; J^{\pi}=1/2^-) =
\Biggl(
-\frac{\nabla^2}{2\mu} + V_{\rm C}^{\rm eff}(\vec r)
\Biggr)
R(\vec r, t-t_0 ~; J^{\pi}=1/2^-)~.
\label{pot}
\end{eqnarray}
Eq. (\ref{pot}) is the time-dependent Schr\"{o}dinger-type equation for the R-correlator,
from which the s-wave meson-baryon effective central potentials are extracted. 

\section{Numerical setup}
In order to calculate the meson-baryon potentials in 2+1 flavor full QCD,
we have utilized the gauge configurations of JLDG(Japan Lattice Data Grid)/
ILDG(International Lattice Data Grid)
generated by PACS-CS Collaboration on a $32^3 \times 64$ 
lattice~\cite{JLDG}. 
The renormalization group improved Iwasaki gauge action
and non-perturbatively $O(a)$ improved Wilson quark action are used 
at $\beta=1.90$, which corresponds to the lattice spacing $a=0.09$ fm
determined from $\pi$, $K$ and $\Omega$ masses.
The physical size of the lattice is about (2.9 fm)$^3$ and the 
the hopping parameters are taken to be 
$\kappa_u=\kappa_d=0.1370$
and $\kappa_s=0.1364$.

In the present simulation, we adopt the spatial wall source located at $t_0$ with
the Dirichlet boundary condition at time slice $t=t_0+32$ in the temporal direction and
the periodic boundary condition in each spatial direction.
The Coulomb gauge fixing is employed at $t=t_0$.
The number of gauge configurations used in the simulation is 399.
With this setup, we obtain $m_{\pi} = 705(2)$, $m_{K}= 793(2)$, 
$M_N= 1590(8)$ and $M_{\Sigma}= 1665(7)$ MeV.

\section{Numerical results and discussion}

\begin{figure}[htb]
\begin{center}
\includegraphics[width=0.45\textwidth]{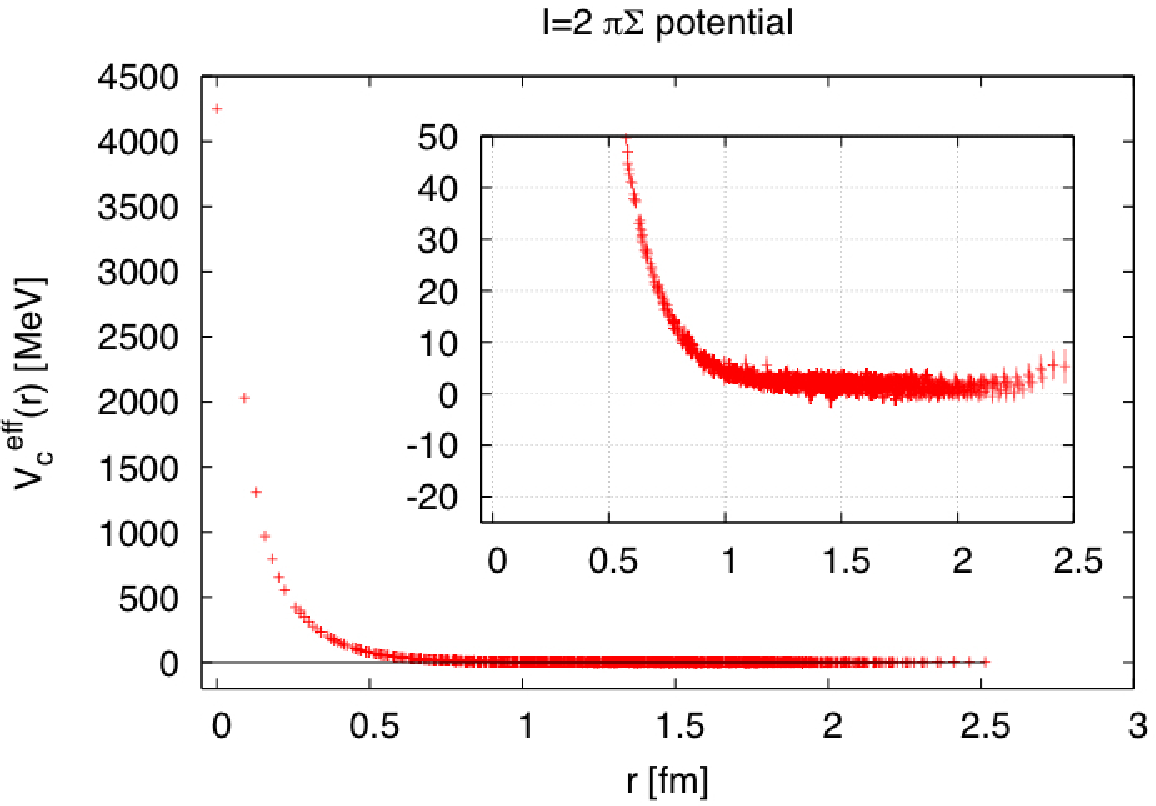}
\put(-155,100){(a)}\hspace{0.5cm}
\includegraphics[width=0.45\textwidth]{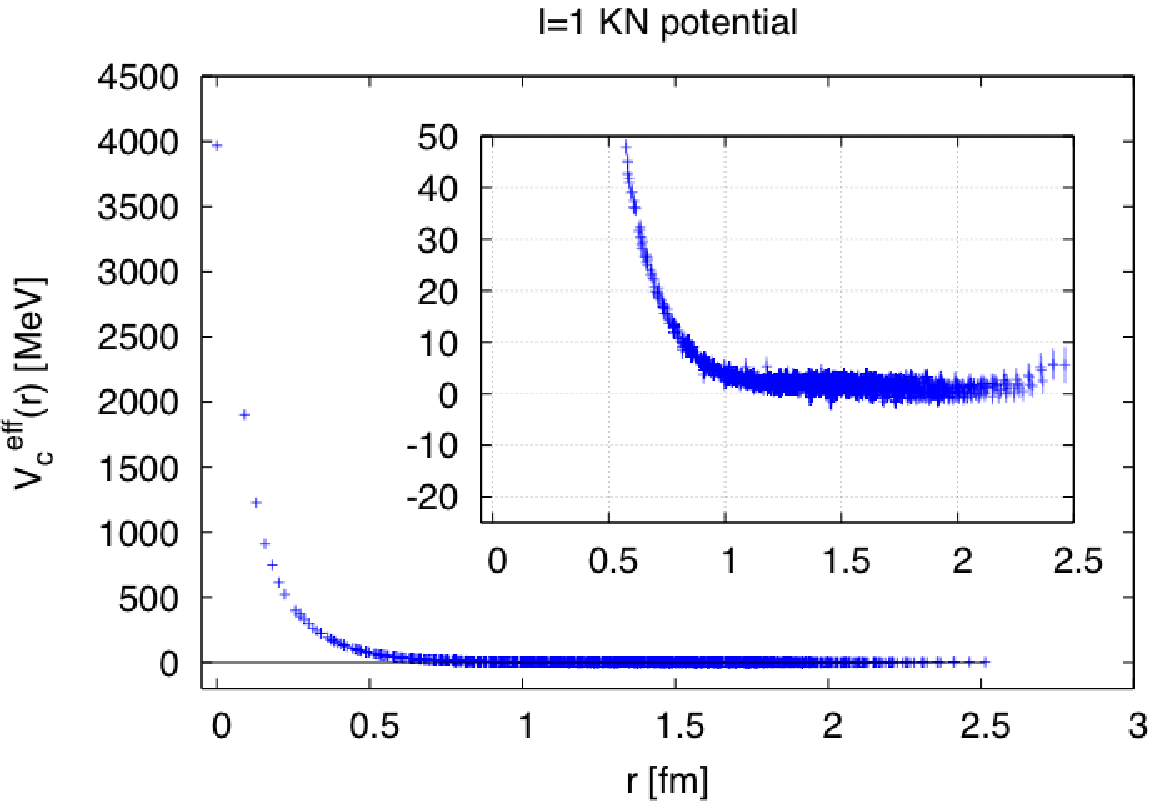}
\put(-155,100){(b)}
\caption{
The s-wave potential of (a) $I=2$ $\pi \Sigma$ 
and (b) $I=1$ $KN$ states.
}
\label{fig1}       
\end{center}
\end{figure}

By using Eq. (\ref{pot}), we obtain the s-wave $I=2$ $\pi \Sigma$ and $I=1$ $KN$
effective central potentials, 
which are shown in Figs.~\ref{fig1}(a) and (b) at time slice $t-t_0=13$.
We observe the repulsive core at short distance ($r<0.5$ fm) in both channels.
The strong repulsions near origin can be expected
by the quark Pauli blocking effects.
This was first pointed out by Machida and Namiki~\cite{MN:65} for the
meson-baryon systems.
In the $I=1$ $KN$ ($K^+ p$) state
whose configuration is 
\begin{equation}
K^+ p \sim (u \bar s)\ (uud)~,
\end{equation}
one of u-quarks can not be in the S-state.
The strong repulsion at short distance in both channels found in our simulations 
suggest a manifestation of the quark Pauli blocking effects.

\begin{figure}[hptb]
\begin{center}
\includegraphics[width=0.45\textwidth]{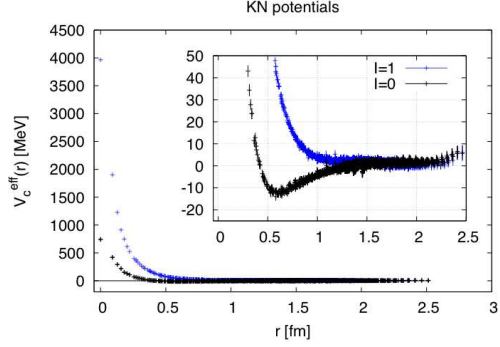}
\caption{
The $I=1$ (blue) and $I=0$ (black) $KN$ potentials.
Short range repulsion in $I=1$ is much stronger than that in $I=0$ 
as predicted by Pauli blocking effects.
}
\label{fig2}       
\end{center}
\end{figure}

The repulsive interactions in the s-wave $I=1$ $KN$ state can be expected
much stronger than that of the $I=0$ $KN$ state
due to the quark Pauli blocking effects.
To investigate whether this expectation is correct or not,
we also calculate the
s-wave $I=0$ $KN$ potential.
As shown in Fig.~\ref{fig2}, 
the repulsion at short distance for the $KN$ potential 
becomes significantly smaller in the $I=0$ channel than $I=1$ channel. 
This again confirms the expectation from  the quark Pauli blocking effects. 
In addition,
we observe the attractive well in the 
mid range ($0.4<r<1.2$ fm) in the $I=0$ $KN$ channel.
In the constituent quark model of hadrons~\cite{Barnes:94},
similar short range repulsion in $KN$ system has been predicted, while
the attraction has not been found.

\begin{figure}[htb]
\begin{center}
\includegraphics[width=0.45\textwidth]{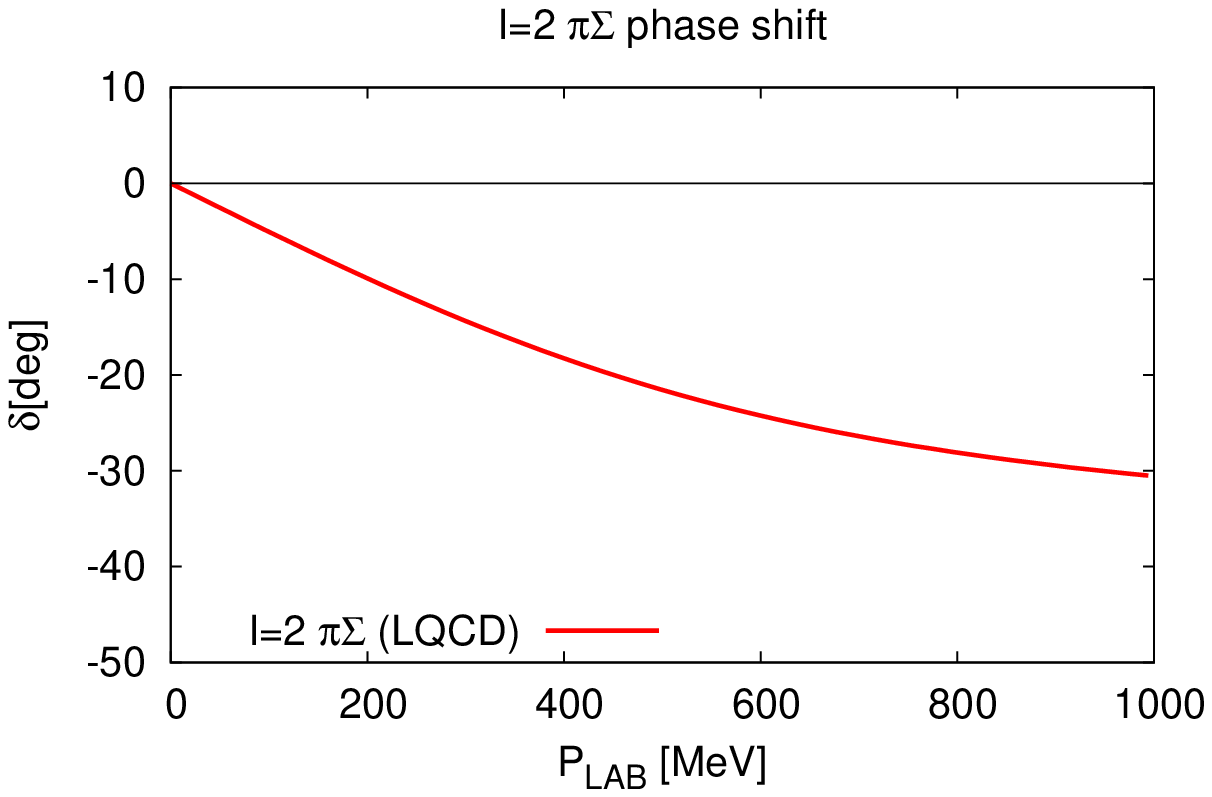}
\put(-155,105){(a)}\hspace{0.5cm}
\includegraphics[width=0.45\textwidth]{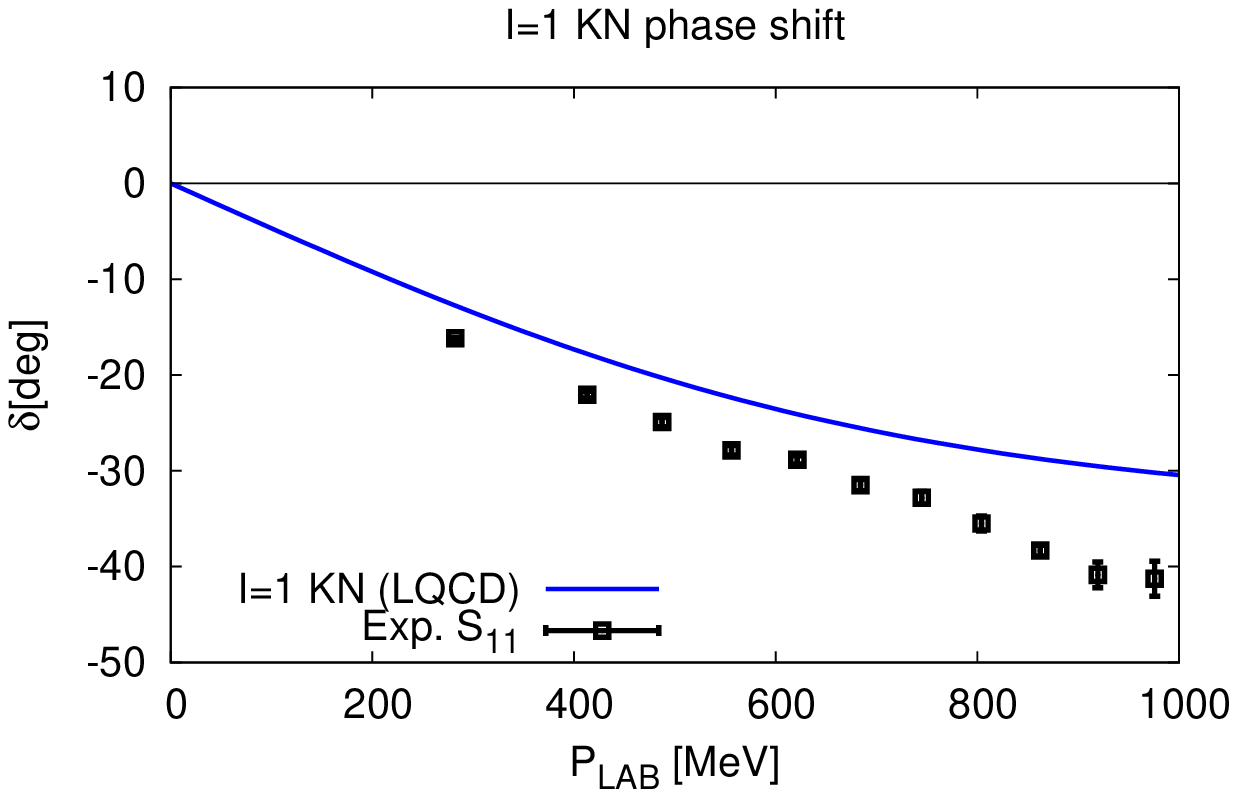}
\put(-155,105){(b)}
\caption{
The s-wave phase shifts of (a) the $I=2$ $\pi \Sigma$ 
and (b) the $I=1$ $KN$ scatterings together with the experimental data~\cite{Hashimoto:84}.
}
\label{fig3}       
\end{center}
\end{figure}

By using the potentials which fit the lattice data in Fig.~\ref{fig1}(a) and (b),
we can calculate observable such as the scattering phase shifts.
Fig.~\ref{fig3} shows such  phase shifts of the $I=2$ $\pi \Sigma$
and $I=1$ $KN$ scatterings together with the experimental data 
as functions of the laboratory momentum of the mesons.
Although the hadron masses are  heavy in the present simulation,
qualitative behavior of the phase shifts
in $I=1$ $KN$ channel is consistent
with the experimental data. 
Simulations along this line with lighter quark masses will eventually
lead to a definite conclusion of the $I=2$ $\pi \Sigma$ scattering.

\section{Summary}
We have performed the 2+1 flavor full lattice QCD simulation to
investigate the $I=2$ $\pi \Sigma$ interaction, which is relevant to determine
the structure of the $\Lambda(1405)$.
The s-wave $I=2$ $\pi \Sigma$ and $I=1$ $KN$ potentials 
are extracted from the NBS wave functions
for relatively heavy quark mass corresponding to $m_{\pi}=700$ MeV.
Potentials in both channels reveal short range repulsions.
From these potentials, the s-wave scattering phase shifts are calculated
and compared with the existing experimental data in $I=1$ $KN$ channel. 
Although the quark mass is heavy in the present simulation, 
the results indicate that our method is promising for future 
quantitative studies of the $\pi \Sigma$ interactions
at lighter quark masses.

\section*{Acknowledgements}
The authors thank PACS-CS Collaboration and ILDG/JLDG~\cite{web}
for providing the gauge configuration.
The calculation were performed mainly 
by using the NEC-SX9 and SX8R at RCNP in Osaka University
and SR16000 at YITP in Kyoto University.
This work is supported in part by 
Grant-in-Aid of JSPS (No. 22-8687),
the  Grant-in-Aid  of  MEXT (No. 20340047),
the Grant-in-Aid for Scientific Research on Innovative Areas
(Nos.~2004: 20105001, 20105003)
and the Large Scale Simulation Program No.09-23(FY2009) 
of High Energy Accelerator Research Organization (KEK).


\end{document}